\documentclass{aa}
\usepackage{graphicx,times,amssymb} \usepackage[english]{babel}
\renewcommand{\d}{\mathrm{d}}
\def\be{\begin{equation}} \def\ee{\end{equation}}
\def\bea{\begin{eqnarray}} \def\eea{\end{eqnarray}}
  
  \def\e{et~al.}

\def\HI{\ion{H}{i}}
\def\CIV{\ion{C}{iv}}
\def\FeII{\ion{Fe}{ii}} 
\def\MgII{\ion{Mg}{ii}}
\def\MgIIFeII{\ion{Mg}{ii}/\ion{Fe}{ii}}
\newcommand{\kms}{\hbox{km s$^{-1}$}}

\begin{document}

\title{Investigating lensing by absorbers in the 2dF-Quasar survey}

\author{Brice M\'enard\inst{1,2,3}\thanks{e-mail:menard@mpa-garching.mpg.de} 
\& C\'eline P\'eroux\inst{4}\thanks{Marie Curie Fellow}} 
\institute{ 
$^1$Max-Planck-Institut f\"ur Astrophysik, 
P.O.~Box 1317, D--85741 Garching, Germany\\ 
$^2$Institut d'Astrophysique de Paris, 
98 bis Bld Arago, F--75014 Paris, France\\
$^3$Shanghai Astronomical Observatory, CAS, 80 Nandan Road, Shanghai 200030, China\\
$^4$Osservatorio Astronomico di Trieste, Via Tiepolo n.11, I--34131
Trieste, Italy}

\date{\today} \authorrunning{M\'enard \& P\'eroux}
\titlerunning{Lensing by absorbers in the 2dF-quasar survey}

\abstract {We use the first data release of the 2-degree Field Quasar survey
  to investigate the effect of gravitational magnification by foreground
  absorbing systems on background quasars. We select two populations of
  quasars from this sample: one with strong \MgIIFeII\ absorbers and one
  without.  The selection is done in such a way that the two populations have
  the same redshift distribution and the absorber detection procedure discards
  possible biases with quasar magnitude. We then compare their magnitude
  distributions and find a relative excess of bright quasars with absorbers.
  This effect is detected at the 2.4, 3.7 and 4.4$\sigma$ levels in u-, b- and
  r-bands.  Various explanations of the observed phenomenon are considered and
  several lines of evidence point towards gravitational lensing causing some
  of the differences observed in the magnitude distributions. We note that
  physical quasar-absorber associations may contribute to some extent to the
  observed correlations for low quasar-absorber velocity differences.  We
  discuss the implications of these findings and propose future work which
  will allow us to strengthen and extend the results presented here.

}

\maketitle
\section{Introduction}

Luminous quasi-stellar objects, hereafter quasars, permit the detection of
arbitrarily faint galactic systems through absorp\-tion in their spectra. These
absorbers are believed to trace various types of systems (disks, dwarfs, low
surface brightness galaxies) and/or different regions of galaxies (the
innermost part, outflows).  Detecting such a system in the spectrum of a
quasar indicates the presence of a matter overdensity along its line-of-sight
and such a concentration of matter may well act as a lens on the background
quasar.  This scenario is then expected to affect the magnitude distribution
of a quasar population showing absorption systems (Bartelmann \& Loeb 1995,
Pei 1995, Perna et al.  1997, Smette et al.  1997).
\newline If some absorbers, for example with high \HI\ column densities, trace
distant galaxies, the associated lensing effects, for impact parameters
greater than $\sim10$ kpc, will likely modify the flux of the source without
producing new additional images.  Since the intrinsic luminosity of a given
quasar is not known {\it a-priori}, such an effect cannot be detected for an
individual object but requires a statistical analysis performed with a large
homogeneous
sample.\\

For a magnitude limited sample of quasars, two effects of gravitational
magnification come into play: the flux received from distant sources is
boosted, increasing the probability of observing quasars behind absorbers,
while the solid angle behind absorbers is gravitationally enlarged, which then
lowers the density of background quasars.  The net result of these competing
effects (an increase or decrease of the number counts of quasars with an
absorber) depends on whether the loss of sources due to dilution is balanced
by the gain of sources due to flux magnification (Narayan 1989).  Sources with
flat luminosity functions, like faint quasars, are depleted by magnification
while the number density of sources with steep luminosity functions, like
bright quasars, is increased. This effect is called the \emph{magnification
  bias} (e.g. Schneider \e\ 1992).

Attempts to quantify this phenomenon have been made in the past through both
theoretical modeling and direct observations.  Pei (1995) estimated the
effects of gravitational lensing by cosmologically distributed dark matter
halos on the quasar luminosity function; Bartelmann \& Loeb (1995) and Smette
et al. (1997) showed how the statistics of Damped Ly-$\alpha$ systems are
affected by lensing; Perna et al. (1997) estimated that for bright quasars
changes in magnitudes due to gravitational lensing by spiral galaxies are
stronger than obscuration effects which give rise to the opposite trend.
More recently, Maller \e\ (2002) have shown that the Sloan Digital Sky Survey
(SDSS; York \e\ 2000) combined with future space-based missions (such as the
Galaxy Evolution Explorer satellite; Milliard \e\ 2001) will provide the
necessary data to constrain the mass to gas ratio of certain types of
absorbers by using the gravitational lensing effects they produce on
background quasars.

On the observational side, one approach consists of fitting lensing models to
the redshift evolution of quasar absorbers to disentangle intrinsic evolution
from gravitational lensing (Thomas \& Webster 1990; Steidel \& Sargent 1992,
Borgeest \& Mehlert 1993).  All these studies found the effect of lensing to
be small for the range of redshifts and equivalent widths they used.  Another
approach first suggested by York \e\ (1991) is to divide quasar spectra into a
bright and a faint sample in order to determine the incidence of absorbers in
each sample separately. In their study, York \e\ found no evidence for
gravitational lensing, except perhaps towards higher redshifts (z$\sim$3).
Vanden Berk \e\ (1996) extended the analysis to a larger sample of quasar
spectra compiled from the literature.  They found an excess of \CIV\ absorbers
in luminous quasars, as would be expected from a gravitational lensing effect,
but did not find a similar trend in the available \MgII\ sample.
Recently, Le Brun et al. (2000) used a sample of 7 Damped Ly-$\alpha$ systems
for which they identified the absorbing galaxies, measured the impact
parameters, and derived the upper limit of 0.3 magnitude for the amplification
factor.

Apart from the latter one, all these above mentioned studies used
  relatively weak absorber samples: considering \MgII , most of the
  corresponding systems they used have an equivalent width
  ($0.2\lesssim\,$W$_0\,(2796\, $\AA $)\lesssim 1$ \AA ).  As a result, the
  authors looked for the cumulative - and likely weak - lensing effects due to
  multiple absorbing systems along the line-of-sight of quasars.  Here we
  adopt another strategy for unveiling the effects of gravitational lensing by
  absorbers: we focus on the strongest systems and look for their
  magnification bias.  Indeed, despite the fact that
  such systems are rarer, some of them may actually probe the inner part of
  galactic halos and might therefore favor observable lensing effects.
  Being able to measure the associated magnification bias might be of great
  interest since it allows us to probe the dark matter distribution of distant
  baryonic systems seen in absorption.  For a given distribution of impact
  parameters and for a given dark matter profile, an accurate measurement can
  then lead to some constrains on the average \emph{mass} of these absorber
  systems.

  In order to detect this effect, we use here the first release of the 2dF
  quasar survey (2QZ), i.e. a large and homogeneous sample of quasar spectra,
  to compare the magnitude distribution of quasars with a strong \MgII\ absorber
  ($1.3\lesssim\,$W$_0\,(2796+2803\, $\AA$)\lesssim 9.0$ \AA ) to that of a
  reference population of quasars for which no such absorption was found.

This paper is organized as follows: in Sect. 2 we present the data and detail
how we selected two unbiased samples of quasars, with and without absorbers.
We then compare their magnitude distribution in Sect. 3 and find a significant
difference between them. In Sect. 4 we review the effects expected from the
presence of a system along a quasar line-of-sight and find gravitational
lensing and physical associations to be the most likely explanations for the
trend we observe.  Finally, we discuss the implications of the results
obtained and propose further extensions of this work to the larger sets of
data which will become available soon.

\section{The Data}\label{data}

Until recently only a few hundred quasar spectra were available for the
detection of quasar absorbers. With the help of multi-fiber spectroscopy,
surveys of thousands of quasars have become publicly available. A pioneer
experiment in this area, the 2 degree field quasar redshift survey (2QZ) has
already acquired more than $20,000$ quasar spectra (Boyle \e\ 2001; Croom \e\ 
2001a; Hoyle \e\ 2002) and provides an unprecedented source for statistical
studies of quasars in a homogeneous sample.

\subsection{The 2dF QSO Redshift Survey and associated quasar absorbers}

The first data release of the 2QZ contains over ten thousand low resolution
quasar spectra, taken with the 2-degree Field instrument at the
Anglo-Australian Telescope in the range 3700 - 7500 \AA .  Quasar candidates
with $18.25<b_J<20.85$ were selected from a single homogeneous color-based
catalogue from APM (Automatic Plate Measuring) measurements of UK Schmidt
photographic material.  Note that the corresponding u- and r-band quasar
catalogues are therefore not complete.
\newline In our study we will use the absorber catalogue compiled by
Outram \e\ (2001). They examined visually the highest signal-to-noise ratio
spectra in order to identify heavy element absorbers (hereafter, ``metal''
absorbers). Their aim was to compile a list of Damped Lyman $\alpha$ (DLA)
candidates for further investigation.  Starting from the 10k catalogue, the
sub-sample used to optimize this ``metal'' search is defined according to the
following criteria:
\begin{enumerate}
\item the sample is limited to quasars with $z_{\rm em}>0.5$;
\item only spectra with signal-to-noise ratio (S/N) greater than 15 in
the range 4000-5000\AA\ were selected. This S/N estimate is based on
spectral variances measured from the 200 fibers in each APM field;
\item quasars exhibiting broad absorption lines or high ionization systems
  with $z_{\rm abs}\sim z_{\rm em}$ are excluded from the analysis.
\end{enumerate}

This selection reduces the 10k release to a sample of 1264 quasars which
were {\em visually} inspected for the detection of absorption lines.
In order to avoid false detections, they require the presence of at
least two absorber species at the same redshift.
We refer the reader to Outram \e\ (2001) for the details of the absorption
line detection technique.

In this subsample of quasars, 129 spectra contain at least one absorption
system and there remain 1135 quasars for which no absorption line was
detected.  In order to compare the magnitude distribution of these two
samples, the corresponding quasars must be selected in a way that does not
introduce any magnitude bias.  The rest of the section describes the details
of this selection.

\subsubsection{The Absorber Population}

In order to optimize the detection of statistical lensing effects, a large
sample of strong absorbers is required. In the 2QZ the most commonly detected
absorbers are \MgII/\FeII\ systems. Indeed, \MgII\ absorp\-tion is a dou\-blet,
allowing for robust identification, and \FeII\ lines present a variety of rest
wavelengths and oscillator strengths, which eases \FeII\ identification in a
given wavelength range.  The \MgII/\FeII\ absorbers identified by Outram \e\ 
(2001) are strong systems with rest equivalent width of the \MgII\ doublets
ranging from 1.3 to 9.0 \AA\ (see Table \ref{table_details}). In the
following, all equivalent widths will refer to the one of the \MgII\ doublet,
as given by Outram \e\ (2001).
This preselection reduces the sample introduced above down to 109 quasars. The
corresponding list is given in Appendix A.  For our analysis, we do not take
into account the small number of quasars which are not detected in the r-band:
one quasar with an absorber (flagged R1 in Table \ref{table_details}) and
2\% of the reference quasars, in order to analyze well-defined samples in
every magnitude band.  Lastly, since our goal is to look for gravitational
lensing effects, we also exclude one system (flagged R2 in the Table) for
which the absorber escape velocity is smaller than $3\,000$ km/s which
indicates that the absorber is very likely physically associated to the
quasar.  In the six cases where two absorbers are detected in a single quasar
spectrum (A flags), we consider only the one having the largest equivalent
width.  (Note that, given the small number of quasars concerned, using other
choices do not significantly affect the results of this study).  The resulting
set of quasars with absorbers contains 108 objects and the reference
population 1114. The selection steps are summarized in Table \ref{table_summary}.\\

As mentioned in the introduction, the amplitude of gravitational lensing
effects by absorbers is expected to be highest when absorbing systems probe
the inner part of galactic halos. Such a situation is likely to occur when
quasar spectra show strong \MgII\ lines or a Damped Lyman-$\alpha$ absorber at
z$\gtrsim$0.5 (Steidel 1995, Le Brun et al. 2000). Several theoretical
studies have investigated the corresponding lensing effects (Pei 1995,
Bartelmann \& Loeb 1995, Perna et al. 1997, Smette et al. 1997).
In the present study, it is important to note that our absorber
sample is actually expected to present such properties: strong \MgIIFeII\ 
systems have proven to be excellent tracers of DLAs.  Indeed, although a
number of DLAs are known at high-redshifts (Storrie-Lombardi \e\ 1996;
Storrie-Lombardi \& Wolfe 2000; P\'eroux \e\ 2001) due to their signature in
optical spectra, discovering these systems at lower redshifts is more
challenging since the observed wavelengths of DLAs are
shifted to the ultraviolet, requiring space-based observations.  An attempt to
overcome this drawback was first proposed by Rao, Turnshek \& Briggs (1995)
who suggested the use of \MgII\ absorbers as tracers of DLA candidates. Their
method is based on observational evidence which indicates that DLAs are always
associated with a \MgII\ system.  Extensions of this work (Le Brun, Vitton \&
Milliard 1998; Rao \& Turnshek 2000; Nestor, Rao \& Turnshek 2002) have shown
that strong \MgII\ and \FeII\ absorption systems observed in optical quasar
spectra are very reliable indicators of the presence of DLAs at low-redshifts.
Therefore, our absorber population can be considered as a sample of DLA
candidates.  In such a case, our selection is expected to favour gravitational
lensing effects.

\subsubsection{Redshift Distributions}

\begin{figure}[h]
\begin{center}
  \includegraphics[width=\hsize,height=.3\vsize]{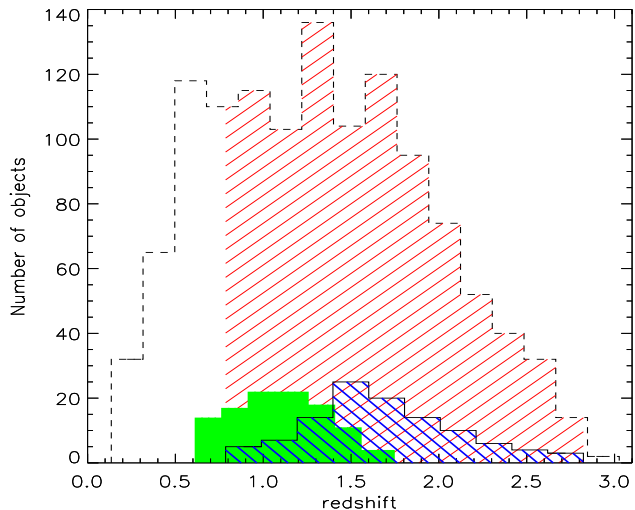}\hfill
  \caption{Redshift distributions of the population of quasars with a strong
    \MgIIFeII\ absorber (solid line), the sample of absorbers (filled
    histogram) and the population of quasars without such absorbers (dashed
    line). For the latter one our analysis considers only quasars whose
    redshifts overlap the ones of quasars with absorbers (red dashed region).
    Since the probability of finding an absorber increases with redshift, the
    sample of quasars with an absorber presents a redshift distribution skewed
    towards high redshifts with respect to the reference population.}
\label{plot_z}
\end{center}
\end{figure}
We present the redshift distributions of the three populations relevant to
this study in Figure \ref{plot_z}: the initial sample of quasars without any
strong absorber (dashed line), the population of quasars with a strong
\MgIIFeII\ absorber (solid line) and the corresponding detected \MgIIFeII\ 
absorbers (filled histogram). Since the probability of finding an absorber
increases as a function of redshift, the redshift distribution of the sample
of quasars with an absorber is skewed towards high redshifts with respect to
the population of quasars without an absorber.  In addition, the apparent
magnitude of a quasar depends on distance as well as intrinsic luminosity.
Therefore, a meaningful comparison of their magnitude distributions requires
the samples to have the same redshift distribution.  Thus, we will work with
subsamples of the population of quasars without absorbers selected such that
the number of objects and the redshift distributions are identical to that of
the quasars with an absorber. This range is represented by the red dashed
region in Figure \ref{plot_z}. Moreover this choice of redshift distribution
automatically rejects the low redshift quasars for which the \FeII\ lines
(2344, 2382, 2587 and 2600 \AA) do not fall in the observed wavelength range.
This set of bootstrapped subsamples constitutes our \emph{reference} sample.

\subsubsection{Signal-to-noise Properties}

The absorber detection efficiency depends on the S/N of the quasar spectra, so
it is necessary to verify how this parameter influences the magnitude
distribution of the two quasar populations. Since each field observed by the
2QZ was exposed approximately for the same length of time (about 55 minutes),
we expect some correlation between the magnitude of the quasars and their S/N.

\begin{figure}[ht]
  \includegraphics[width=\hsize,height=.5\vsize]{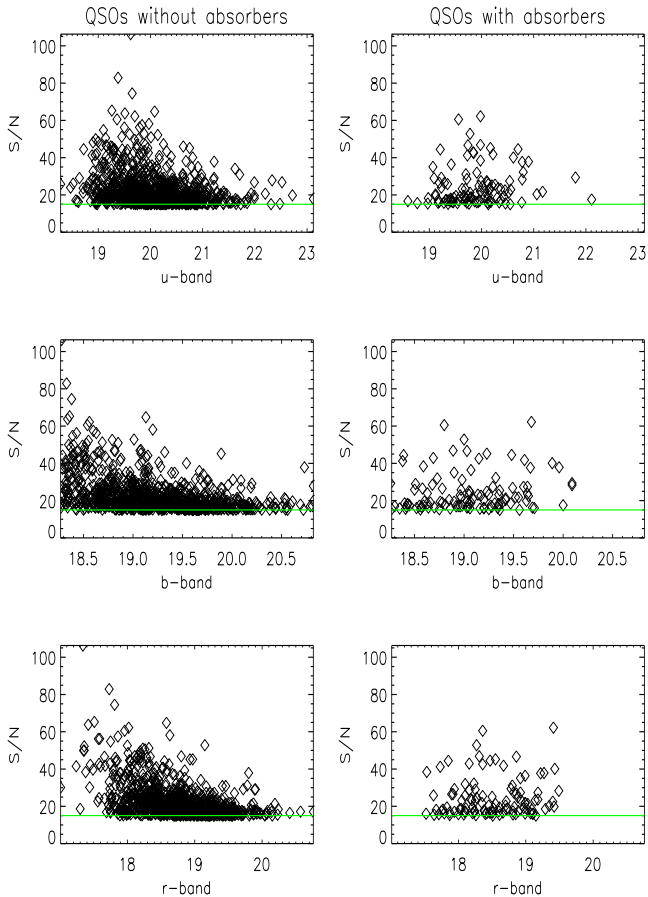}\hfill
  \caption{Magnitude-S/N distributions of the two quasar populations
    in u-, b- and r-bands. The reference population is plotted on the left
    hand side and the population of quasars with a detected absorber on the
    right hand side. A correlation is clearly seen in each band. The S/N
    parameter is the one given by the 2dF, i.e. computed in the range
    4000-5000\AA\ .  The horizontal light-coloured line defines the limit
    S/N$>$15 used by Outram et al. (2001) in order to define the quasar
    sample.}
\label{plot_sn_mag}
\end{figure}

In Fig. \ref{plot_sn_mag}, we plot the 2dF S/N estimate as a function of the
magnitude of the quasars in the u-, b- and r-bands. The right panels show the
population of quasars with a detected absorber and the left ones the reference
population.  Bright quasars tend to have higher S/N as is expected; the
scatter of this correlation is quite large since the S/N not only depends on
magnitude, but also on the sky brightness, the exposure time and the air mass.
Given this correlation, it is important to check that the absorber selection
does not bias the sample towards a preferential range of S/N and hence biases
the magnitudes.
\newline In a spectrum with a resolution FWHM, the minimum
detectable equivalent width with a $n\sigma$ significance is:
\begin{equation}
\label{eq_w}
{\rm W_{min}}=n\times {\rm FWHM} \times {\rm(S/N)^{-1}}\;,
\end{equation}

\begin{figure}[ht]
  \includegraphics[width=\hsize,height=7cm]{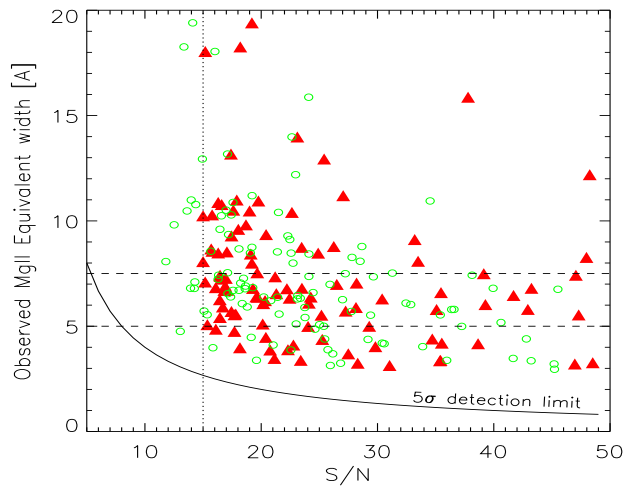}
  \caption{Observed \MgII\ equivalent width W as a function of
    spectrum S/N. The typical uncertainty on W is of the order of 20 per cent.
    The solid line indicates the region above which \MgII\ absorbers are
    detected at more than 5$\sigma$. The triangles show the 2dF S/N estimate
    based in the 4000-5000 \AA\ range and the circles show our estimate using
    the 5000-7000 \AA\ range. The two horizontal lines represent the lower
    limits used in section \ref{detection} to discard possible biases in the
    absorber detection procedure. Note that five systems with S/N$>$50 are
    not displayed for clarity.}
\label{W_0}\end{figure}

\begin{figure*}[t]
  \includegraphics[width=\hsize,height=8cm]{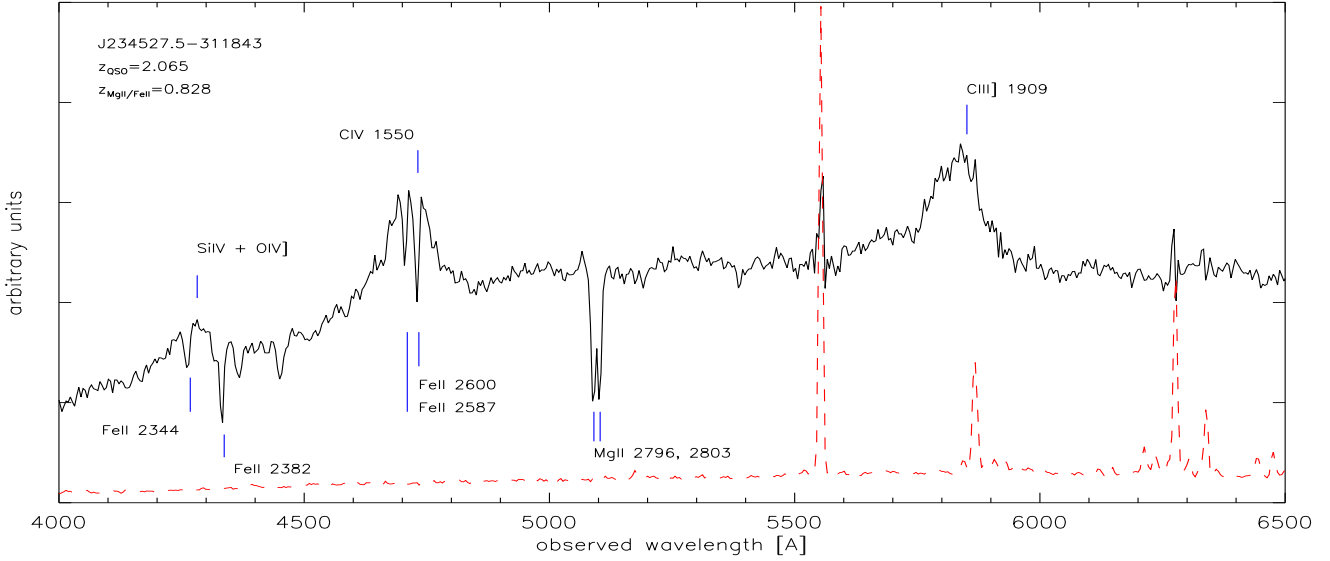}\hfill
\caption{Example of \MgIIFeII\ detection. The spectrum of this quasar at
  $z_{\rm em}=2.06$ clearly shows the \MgII\ doublet as well as the four
  \FeII\ lines expected for an absorber at $z_{\rm abs}=0.82$.  This spectrum
  has a signal-to-noise ratio of 48.  The dashed line is the corresponding sky
  spectrum.}
  \label{plot_spectrum}
\end{figure*}

Given the resolution of 2dF spectra (FWHM = 8\AA), the minimum equivalent
width detectable at 5$\sigma$ for a S/N greater than 15 is
W$_{\rm {min}}=2.6$\AA\ .\\
The visual detection of \MgII/\FeII\ systems is mostly based on the \MgII\ absorption
feature, since the \MgII\ doublet is the more distinct feature (see Figure
\ref{plot_spectrum} for an example).  The equivalent widths of the \MgII\ 
absorbers we use are given by Outram \e\ (2001, see Table 1 of their paper). In
Fig. \ref{W_0} we show the distribution of the observed \MgII\ 
equivalent widths as a function of spectrum's S/N with triangles. 
Following Eq. \ref{eq_w}, the solid line represents the region above which
\MgII\ absorbers can be detected at least at the 5$\sigma$ level.
Since all the absorbers lie above the solid line of this figure, all of them
have a sufficient detection level ($>5\sigma$) to be identified in any of the
quasar spectra of the sample. Therefore, based on the \MgII\ absorption line
detection, our sample is not biased towards any preferential S/N or magnitude
range.
%
The fact that all the points lie well above the $5\sigma$ level depends on the
criteria used by Outram et al. (2001) in their absorber search (for example
the requirement of the presence of \FeII\ lines) and show that only very
robust detections are considered. We have further checked this by visually
inspecting the spectra of the objects listed by Outram et al. and found that
most of the systems show the four Fe II lines (2344, 2382, 2587 and 2600 \AA)
and all of them clearly show at least three of the four lines. The
simultaneous detection of several lines (in addition to the presence of \MgII
) makes the \FeII\ detection quite robust.

Besides, it is worth pointing out that given the rest frame wavelengths of
\MgII\ transitions (2796 and 2803\AA), a number of the corresponding
absorption lines fall in the range 5000-7000\AA\ of the quasar spectra.
However, as mentioned above, the S/N-parameters as determined by the 2QZ team
are measured in the range 4000-5000\AA. Since the S/N of a spectrum is a
function of the observed wavelength, the 2QZ S/N may not be optimal for
testing how the absorber detection technique might bias the luminosity
function of our two quasar populations. In order to obtain a S/N parameter
that more directly describes this selection bias, we have recomputed the S/N
of each quasar in the observed wavelength range 5000-7000\AA\ by using the
ratio between the raw and smoothed spectra. Appropriate masks have been used
at a number of locations where sky lines were not completely subtracted
during data reduction.
\newline The circles in Figure \ref{W_0} show the distribution of the observed
\MgII\ equivalent widths as a function of our spectrum's S/N estimator made in
the range 5000-7000\AA\ .  The fact that all the circles lie also above the
solid line shows that the \MgII\ absorber detection is not biased towards any
preferential S/N in the whole 4000-7000\AA\ range of the spectra.

\section{The effect of absorbers on the quasar magnitude distribution}
\label{detection}

\begin{figure*}[t]
  \includegraphics[width=.33\hsize,height=7.5cm]{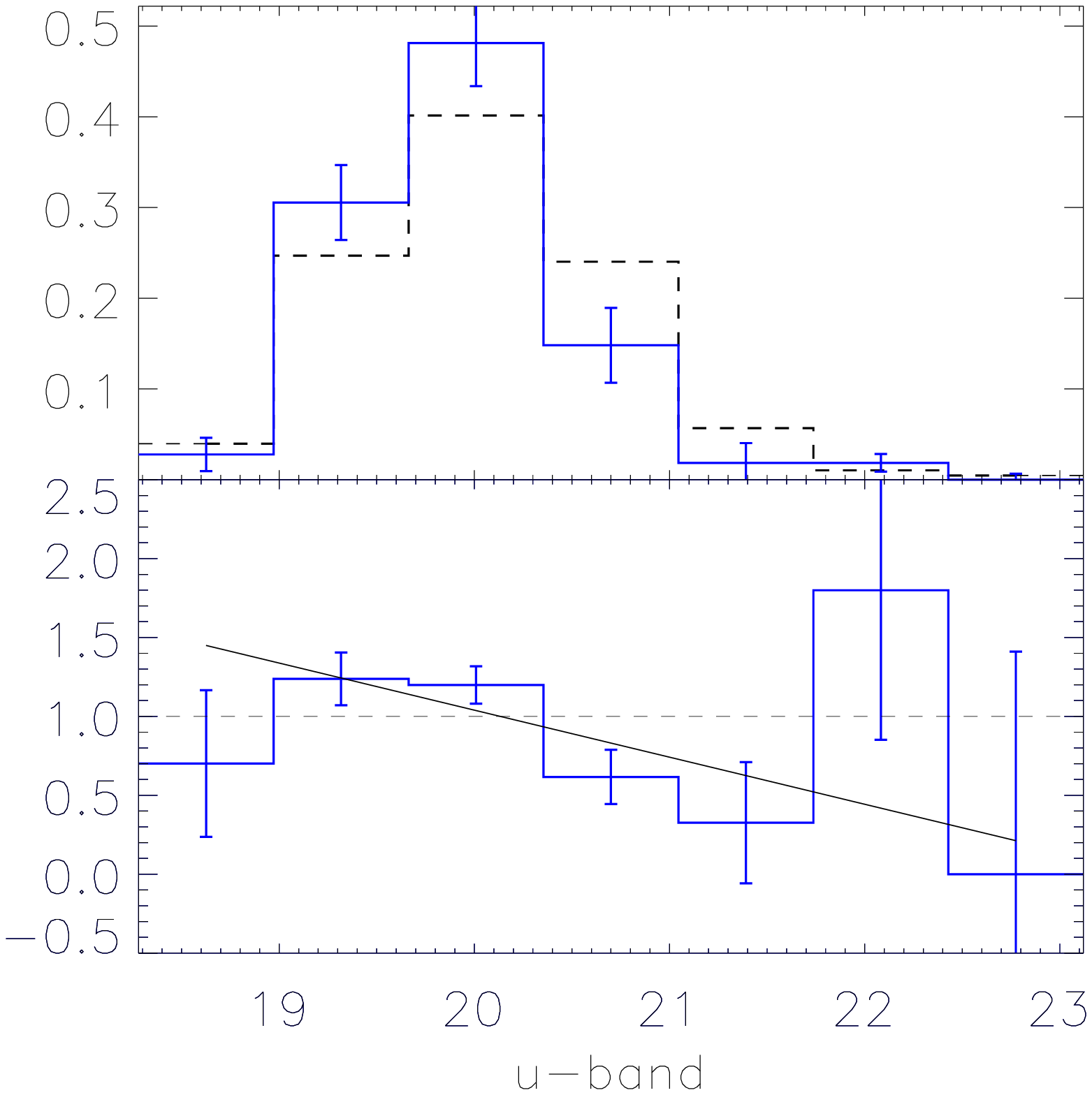}\hfill
  \includegraphics[width=.33\hsize,height=7.5cm]{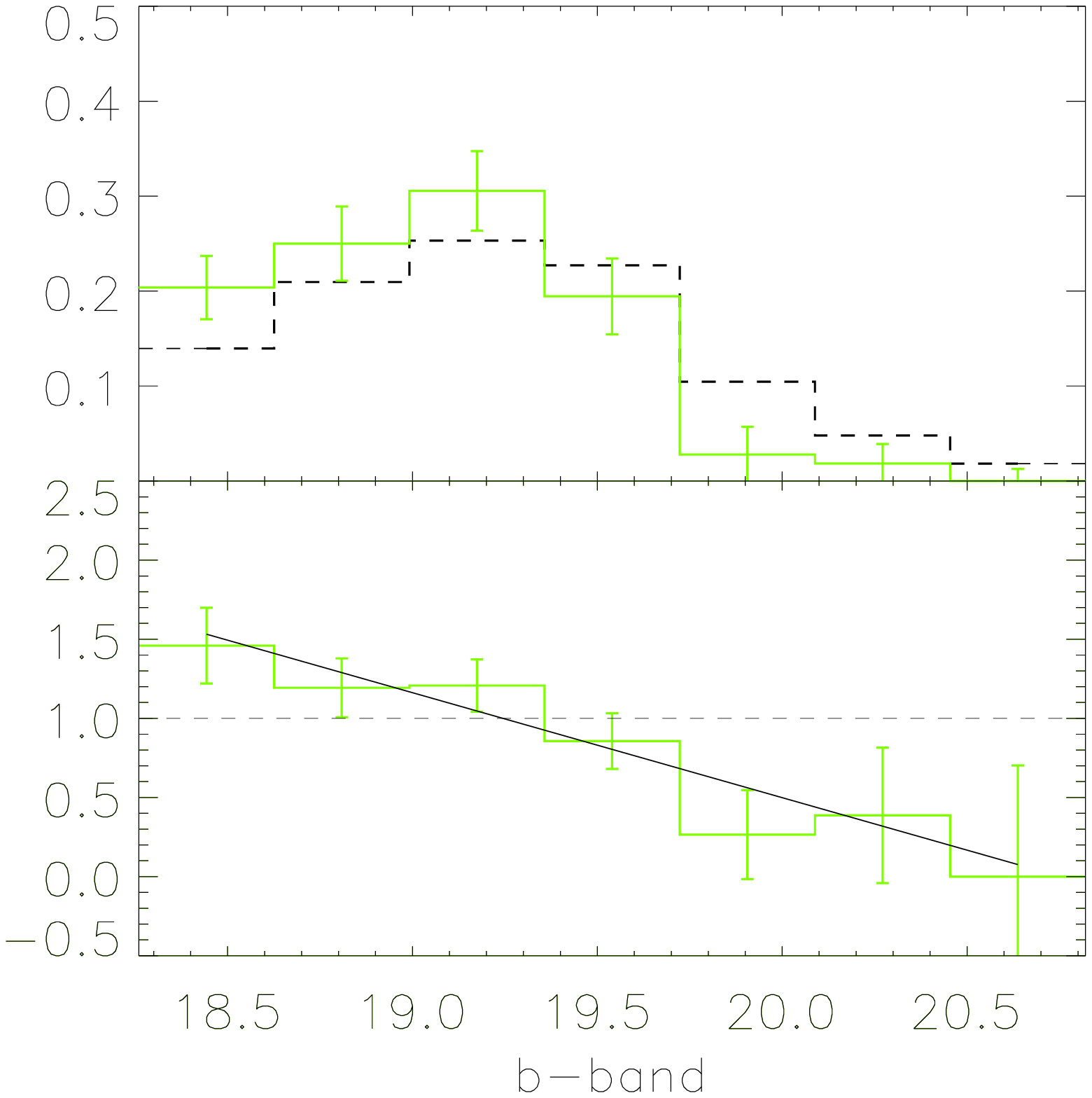}\hfill
  \includegraphics[width=.33\hsize,height=7.5cm]{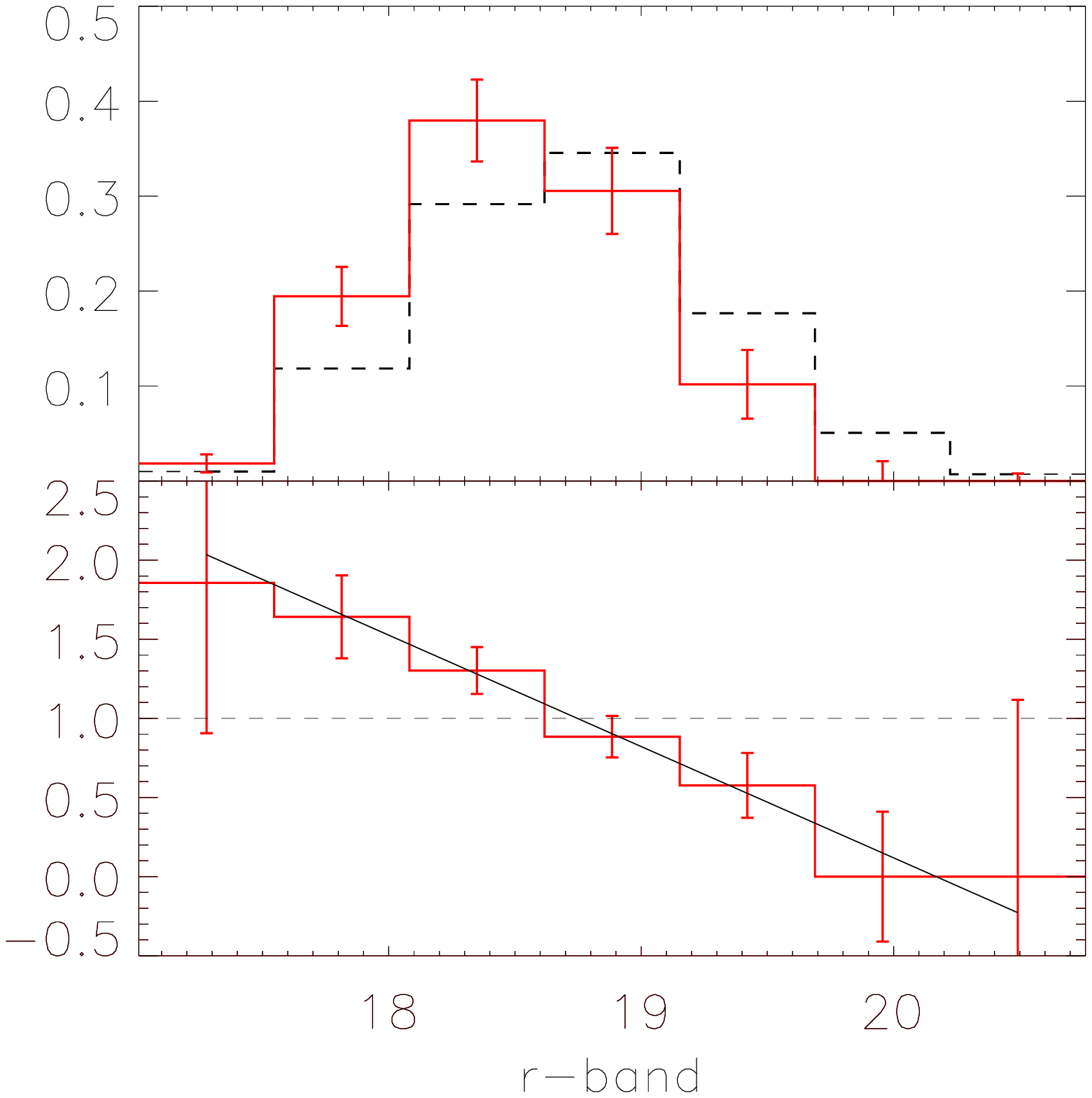}\hfill
  \caption{The top panels present the u-, b- and r-band magnitude
    distributions of the population of quasars with a strong \MgIIFeII\ 
    absorber in solid lines and the quasars without such absorbers in dashed
    lines. The latter is computed from a combination of bootstraps that match
    the redshift distribution of the first. The bottom panels show the
    distribution ratios, as well as noise coming from the finite size of the
    small sample of quasars with an absorber. Each magnitude band clearly
    indicates a tilt between the two magnitude distributions.}
  \label{plot_results_tot}
\end{figure*}
Now we compare the magnitude distributions of the two quasar populations,
knowing that a difference can only be related to the presence of
the absorber along the quasar line-of-sight.\\
We proceed as follows: from the reference quasar population we bootstrap
$10\,000$ subsamples having the same size as the sample with absorbers, i.e.
N$_\mathrm{QSO}=108$ (see Table \ref{table_summary} for the successive steps
of the object selection).  As previously mentioned, the objects are chosen
from the reference sample with the same redshift distribution as the sample
with absorbers.  The combined 10\,000 bootstrap samples without absorbers are
used as a reference sample in what follows.

\begin{figure*}[ht]
  \includegraphics[width=.33\hsize,height=6.5cm]{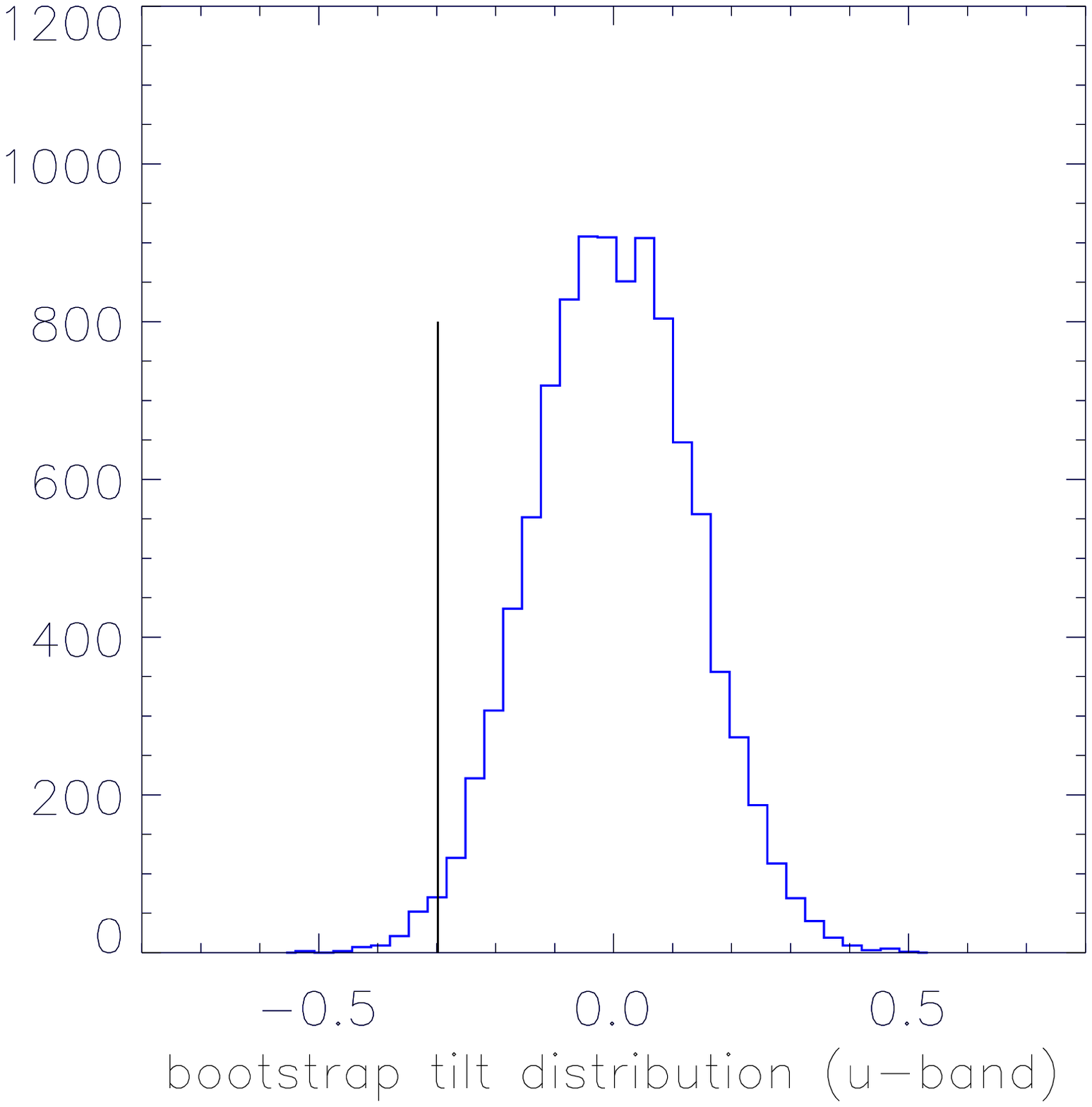}\hfill
  \includegraphics[width=.33\hsize,height=6.5cm]{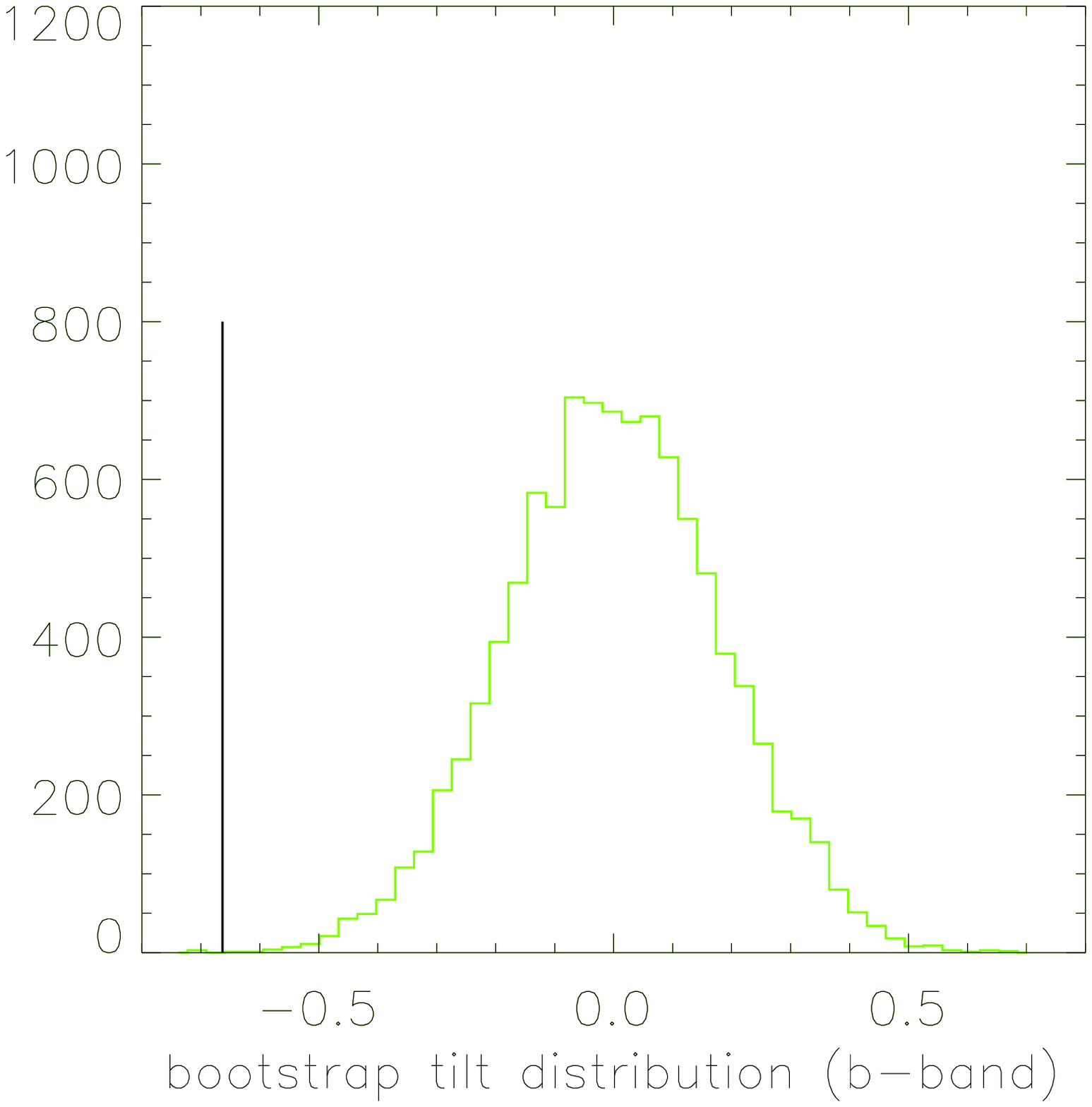}\hfill
  \includegraphics[width=.33\hsize,height=6.5cm]{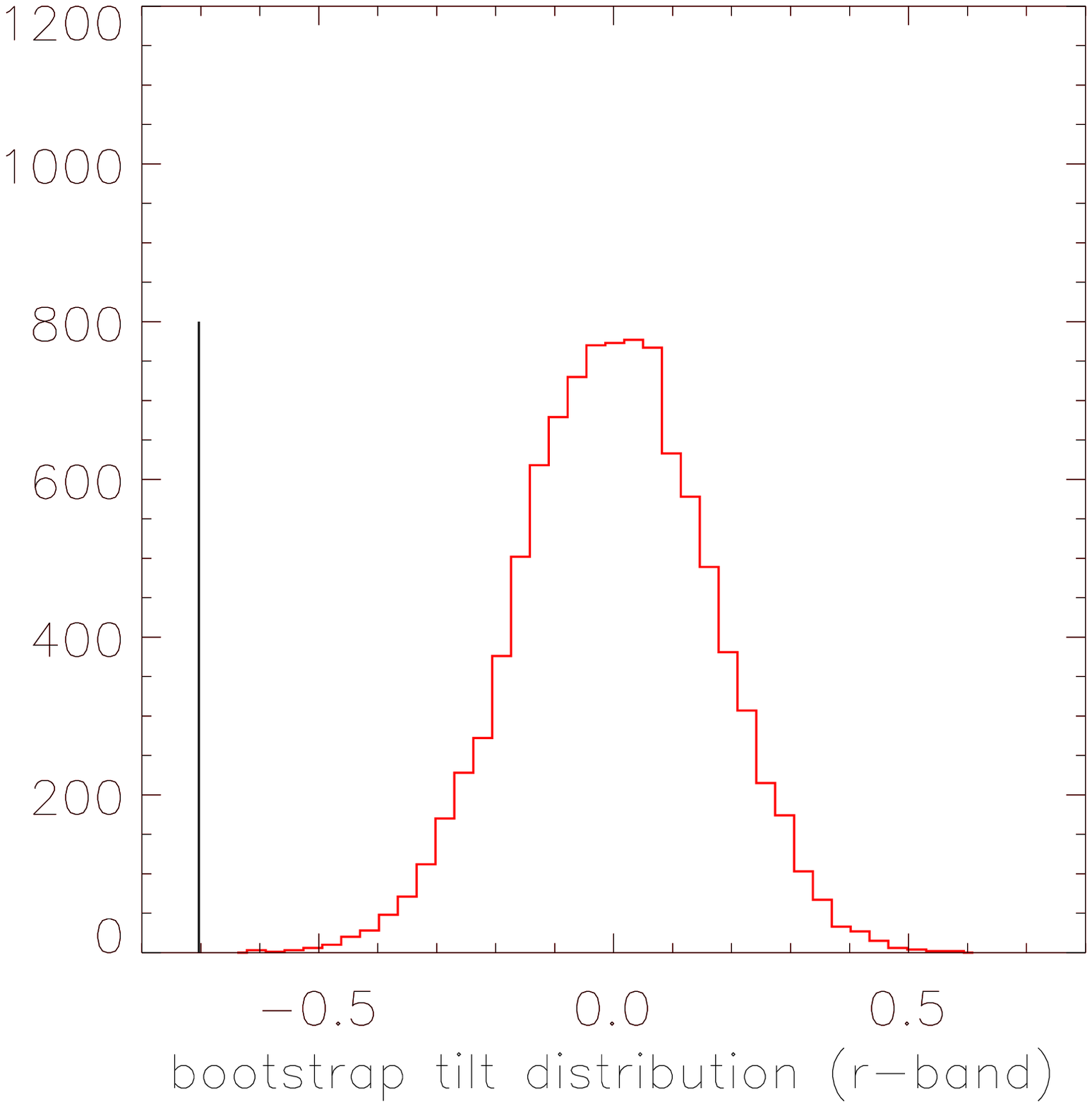}\hfill
  \caption{Distributions of the gradients $\gamma_\mathrm{boot}$ obtained by
  fitting the ratio between the 10\,000 bootstraps subsamples and the
  combined reference sample, in u-, b- and r-bands. The scatter of
  these distributions allows us to define the significance levels of
  the tilt detection for the sample of quasars with an absorber. These
  detections are shown with the vertical lines.}
  \label{plot_results_boot}
\end{figure*}

In the upper panels of Fig. \ref{plot_results_tot} we show the magnitude
distribution of the two quasar samples in the u-, b- and r-bands.  The
solid line represents the magnitude distribution of the quasars with an absorber
and the dashed line shows the combined reference sample. Since these two
populations have different sizes, we plot the fraction of objects per
magnitude bin. The magnitude distribution of quasars with an absorber appears to
be skewed towards bright objects with respect to the reference population.
This effect is observed in each band and suggests an intrinsic difference in
the magnitude distributions. We now show that this difference is significant:
we estimate the Poisson noise associated with the number of quasars with
absorbers per magnitude bin $N_\mathrm{QSO+abs}(m)$ by computing the r.m.s.
deviations in each bin of the bootstrap subsamples introduced above. We
neglect the noise contribution of the reference population
$N_\mathrm{QSO,ref}$, i.e.  the combined bootstraps, since it is a much larger
sample.
For each band, we compute the ratio between the two magnitude distributions:
$N_\mathrm{QSO+abs}(m)/N_\mathrm{QSO,ref}(m)$, normalised by the number of
objects.
The ratios are displayed in the lower panels of Fig. \ref{plot_results_tot}.
Each of them is then fitted by a straight line with gradient $\gamma$.  If
both samples had similar distributions, then the ratio would be unity and
$\gamma$ would be zero. However, in each band, these fits indicate a tilt
between the two magnitude distributions, showing an excess of bright (and/or a
deficit of faint) quasars with an absorber with respect to the reference
population. The corresponding gradients are $\gamma_u=-0.29\pm 0.12$,
$\gamma_b=-0.66 \pm 0.17$ and $\gamma_r=-0.70 \pm 0.15$ in the u, b, and r
bands, respectively.
The significance of these detection is computed by applying the same
analysis to the 10\,000 bootstrapped samples, i.e. we compare each of them to
the combined reference population, by computing the corresponding ratios and
fitting them with straight lines. Fig. \ref{plot_results_boot} shows the
corresponding distribution of gradients $\gamma_\mathrm{boot}$ for each band,
as well as the value found for the sample of quasars with absorbers (vertical
line). Each distribution is well fitted by a Gaussian.
It appears that the magnitude distribution of the quasars with absorber
significantly deviates from the magnitude distributions of 
random reference samples.
Bright quasars with an absorber are in excess compared to the
reference population (and/or faint quasars are in deficit). The tilt between
the two populations is seen at the 2.4, 3.7 and 4.4$\sigma$ level in the u-,
b- and r-bands respectively.  Table \ref{table_gradient} summarizes the
amplitude of these detections. Note that these values are relatively weakly
sensitive to the number of bins used.

Given the fact that we fix the number of quasars in the bootstrap subsamples
to match the number of quasars with an absorber, the only relevant information
in our comparison is the value of the tilt, i.e. the gradient $\gamma$
between the magnitude distributions. The magnitude for which the two
distributions are similar (the zero point) can not be estimated with this
technique and therefore the observed tilt could arise either from an excess of
bright quasars with an absorber, or from a deficit of faint quasars with
absorbers, or from a combination of the two.
This analysis shows, that the magnitude distributions and therefore the number
counts of quasars with such \MgIIFeII\ absorbers are significantly different
from the ones of similar quasars without such absorption lines.  Equivalently,
more absorbers are found in front of bright quasars.

In order to further discard the existence of biases towards any preferential
S/N during the absorber detection technique (which we note was based on a
visual inspection), we have also performed our analysis on two subsamples of
\MgIIFeII\ absorbers having observed equivalent widths $W_{obs}>5.0$ and
$W_{obs}>7.5$ \AA . Indeed, the greater the observed equivalent widths are, the
more robust their detection in spectra are. The corresponding results are
presented in Table \ref{table_gradient}.  As we can see, the signals for the
larger equivalent widths subsamples show gradients that are consistent with
the main sample.  This test strengthens the interpretation that biases in the
absorber detection procedure are not responsible for the differences measured
in the magnitude distributions, and that the relative excess of bright quasars
with an absorber is real.

\begin{table}[ht]
\begin{center}

\begin{tabular}{ll}
  \begin{tabular}{c}
    Main sample:\\
    108 QSO with abs.
    \end{tabular}&  \begin{tabular}{lcc}
                        \hline\hline
                        band    &gradient $\gamma$ &detection level\\
                        \hline 
                        u &-0.29$\pm 0.12$  & 2.4$\sigma$\\
                        b &-0.66$\pm 0.17$  & 3.7$\sigma$\\
                        r &-0.70$\pm 0.15$  & 4.4$\sigma$\\
                        \hline\hline
                      \end{tabular}\\
~\\
\end{tabular}

\begin{tabular}{ll}
  \begin{tabular}{c}
    $\mathrm W_{obs}>5.0$ \AA\ \\
    ~~83 QSO with abs.
    \end{tabular}&  \begin{tabular}{lcc}
                        \hline\hline
                        band    &gradient $\gamma$ &detection level\\
                        \hline 
                        u &-0.35$\pm 0.14$  & 2.3$\sigma$\\
                        b &-0.62$\pm 0.19$  & 3.0$\sigma$\\
                        r &-0.65$\pm 0.17$  & 3.5$\sigma$\\
                        \hline\hline
                      \end{tabular}\\
~\\
\end{tabular}

\begin{tabular}{ll}
  \begin{tabular}{c}
    $\mathrm W_{obs}>7.5$ \AA\ \\
    ~~37 QSO with abs.
    \end{tabular}&  \begin{tabular}{lcc}
                        \hline\hline
                        band    &gradient $\gamma$ &detection level\\
                        \hline 
                        u &-0.28$\pm 0.21$  & 1.2$\sigma$\\
                        b &-0.78$\pm 0.29$  & 2.5$\sigma$\\
                        r &-0.90$\pm 0.25$  & 3.3$\sigma$\\
                        \hline\hline
                      \end{tabular}\\
\end{tabular}

\end{center}
\caption{This table lists the gradient $\gamma$ of the ratio between the
magnitude distribution of the quasars with an absorber and the reference
population, as well as the significance level of a non-constant ratio
($\gamma$=0), for each magnitude band. These significance values come from
the 10\,000 bootstrap subsamples (see Fig \ref{plot_results_boot}).}
\label{table_gradient}
\end{table}

\section{Interpretation}

Given that the selection procedure of the two quasar populations discards
biases with magnitude, the observed differences must be related to the
presence of absorbers.  Indeed, such a structure along the line-of-sight of a
quasar can modify the quasar magnitude in different ways. Firstly, the
presence of dust in the absorber could cause some extinction and redden the
quasar light. Secondly, if the absorber is a galaxy it could also contribute
to the observed luminosity of the quasar. Thirdly, if this absorber system is
sufficiently massive it can gravitationally lens the background quasar and so
modify its magnitude.  Finally, a fraction of the absorbers could be
physically associated with the quasars and therefore correlate with their
physical properties and orientations. In this section, we review these effects
and estimate their impact on the quasar magnitude distribution.

\subsection{Obscuration effects}

Absorber systems are believed to contain dust which produces
extinction effects on the background quasar (Pei \& Fall
1995). This phenomenon is wavelength-dependent: bluer parts of the
spectrum are more affected, rendering the absorption-line sample to
appear redder than the reference one. 

Extinction effects will shift the quasar magnitude distribution to fainter
magnitudes. Given the shape of these distributions in our case (see Fig.
\ref{plot_results_tot}), extinction effects should increase the number of
\emph{faint} quasars and thus tilt the magnitude distribution of the quasars
with an absorber to a positive gradient $\gamma$. In contrast, we detected the
opposite effect in the previous section, i.e. a negative gradient.  This
implies the existence of a phenomenon related to the absorber whose amplitude
is opposite to and dominates over extinction effects.

Besides, we note that since extinction effects are expected to be stronger in
bluer bands, this could well explain the lower amplitude of the tilt observed
in the u-band compared to the r-band.

\subsection{Flux Contribution from the Absorber}

It is believed that the systems responsible for the \MgII/\FeII\ absorptions
observed in the background quasar spectra might be associated with
galaxies.  If these galaxies are bright enough and if the impact parameter is
small enough, they could contribute to the magnitudes measured for the quasars.
The contribution to the ratio between the apparent luminosity distributions of
the quasars with and without absorbers is:
\begin{equation}
R(l_{QSO}^{obs})=\frac{l_{QSO}+l_{absorber}}{l_{QSO}}\,.
\end{equation}
Therefore this effect would increase the relative
number of \emph{faint} quasars with respect to the number of bright
ones. It would therefore give rise to a positive gradient $\gamma$, as
opposed to the negative values observed in the previous section.\\
It is known from deep observations of both DLAs and \MgII\ systems (Le Brun
\e\ 1997; Boiss\'e \e\ 1998; Bergeron \& Boiss\'e 1991) that the luminosity of
these objects is small. Thus, the flux contribution of the absorber to the
quasar luminosity is probably small in general. Moreover, absorbers are
located at various redshifts and have different luminosities. Since these
parameters are uncorrelated with the properties of the background quasar, such
effects should further reduce the absorber flux contribution to the global
magnitude distribution.

\subsection{Gravitational Lensing}

Under the assumption that the absorber systems are intervening galaxies along
the lines-of-sight, a likely explanation for the observed trend in the
magnitude distributions is the \emph{magnification bias} due to gravitational
lensing. Indeed, gravitational magnification has two effects:

\begin{itemize}
\item first, the flux received from background quasars is increased
 by a magnification factor $\mu$ which is related to the
overdensities of matter along their line-of-sight;
\item on the other hand, the solid angle in which sources appear is
  stretched. The probability of observing such quasars is reduced.
\end{itemize}
Additionally, a 'by-pass' effect causes the lines-of-sight towards background
quasars to avoid the central parts of galaxies and reduces their effective
cross-section for absorption.

\begin{figure}[ht]
  \begin{center}
  \includegraphics[width=\hsize,height=.25\vsize]{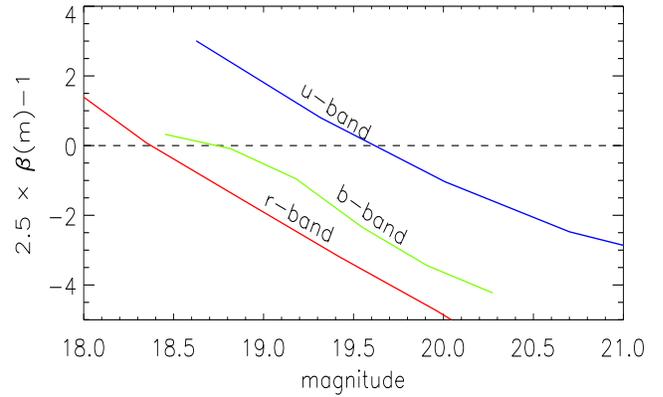}\hfill
  \caption{The value of $2.5\times\beta(m)-1$ is shown as a function of
    magnitude in the u-, b- and r-bands and indicate the relative excess of
    bright quasars with absorbers expected from the magnification bias. Note
    that the three bands span different magnitude ranges.}
\label{plot_nb_counts}
\end{center}
\end{figure}

Considering the first effects, the relative change in the number of quasars
with absorbers depends on the magnification factor $\mu$ and the shape of the
quasar magnitude distribution. Indeed, this effect shifts the magnitude
distribution of the background sources towards brighter values. The steeper
the number counts, the higher the increase.  The number of sources with a
steep luminosity function, like bright quasars, will be increased.
On the other hand if the local number counts decreases as a function of
magnitude, the corresponding number of lensed quasars is reduced.  As we show now,
this effect depends on the magnification factor $\mu$ and the gradient of the
number of sources as a function of magnitude.  Let $\mathrm{n_0}(s)\,\d s$ be
the number of unlensed quasars with a flux in the range
$[s,s+\d s]$ and $\mathrm{n}(s)\,\d s$ the corresponding number of lensed quasars.
Let's write the unlensed source counts as
\begin{eqnarray}
\mathrm{n_0}(s)\,\d s &=& a\,s^{-\beta(s)}\,\d s\,.
\end{eqnarray}
The magnification effect will enlarge the sky solid angle, thus modifying the
source density by a factor $1/\mu$, and at the same time increase their fluxes
by a factor $\mu$. These effects act as follows on the number of lensed
sources:
\begin{eqnarray}
\mathrm{n}(s)\,\d s &=& \mathrm{Prob}(\mu)\times \frac{1}{\mu}\;\mathrm{n_0}\,(\frac{s}{\mu})
\;\frac{\d s}{\mu} \nonumber \\
         &=& \mathrm{Prob}(\mu)\times  
         \mu^{-2}\,a\, \left( \frac{s}{\mu} \right)^{\beta(s/\mu)}\, \d s
\end{eqnarray}
where Prob$(\mu)$ is the probability of having a lens with magnification $\mu$
giving rise to the absorption lines of interest in the quasar spectrum.
This coefficient plays only the role of a normalisation factor.
If $\beta$ does not vary appreciably over the interval $[s,s/\mu]$, which is
well satisfied if $\mu$ departs weakly from unity, then
\begin{eqnarray}
\mathrm{n}(s)   &\approx& \mathrm{Prob}(\mu)\times \mu^{\beta(s)-2}\,n_0(s)
\end{eqnarray}
which can be written as a function of magnitude as
\begin{eqnarray}
\mathrm{n}(m)\,  &\approx& \mathrm{Prob}(\mu)\times \mu^{2.5\;\beta(m)-1}\,\mathrm{n_0}(m)
\label{eq_mag}
\end{eqnarray}
where $\beta(m)=\d \log[\mathrm n_0(m)]\,/\,\d m$.  The final effect, i.e. a
relative excess or deficit of lensed quasars with a magnitude $m$, is then controlled
by the value of $\beta(m)$.  We have computed this quantity using the values
of $\mathrm n_0(m)$ given by the sample of reference (i.e. unlensed) quasars
introduced above. The results are plotted in in Fig.  \ref{plot_nb_counts}. As
we can see, it indicates that the magnification bias would give rise to a
relative excess of bright quasars with absorbers whereas faint quasars are
depleted.  Here we note also that the magnification bias \emph{is} chromatic
since the $\beta$ value depends on the observed wavelength. In some cases,
such an effect can thus introduce some correlations between the absorbers and
the quasar magnitudes, colours, spectral index for example.

The additional effect mentioned above, the by-pass effect, is detailed in Smette
et al. (1997) and Bartelmann \& Loeb (1995). It systematically reduces the
number of observable high-equivalent widths absorption lines for impact
parameters smaller than the Einstein radius of the lens and is independent on
the magnitude of the background sources.\\

\begin{figure*}[ht]
  \includegraphics[width=.33\hsize]{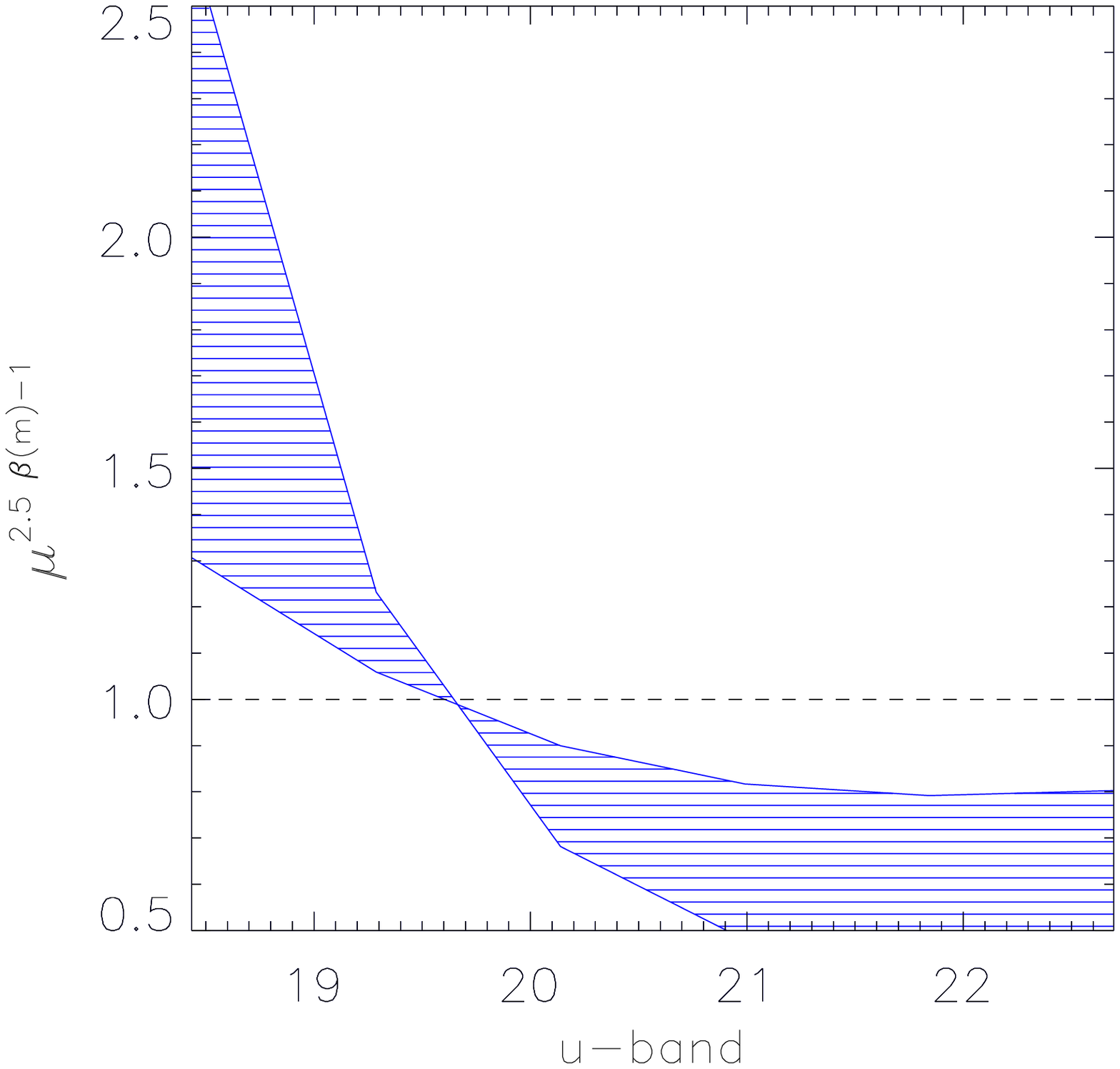}\hfill
  \includegraphics[width=.33\hsize]{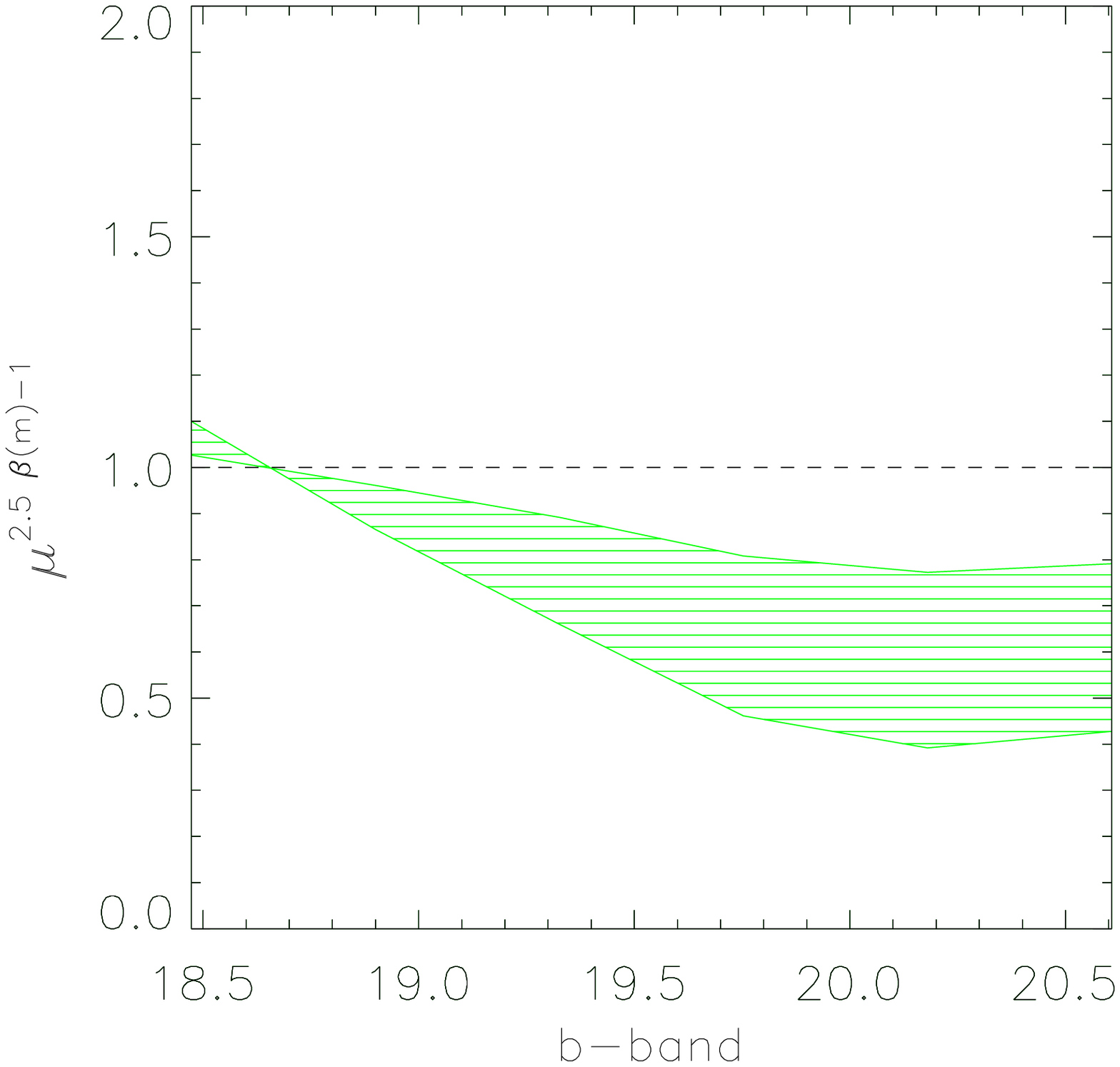}\hfill
  \includegraphics[width=.33\hsize]{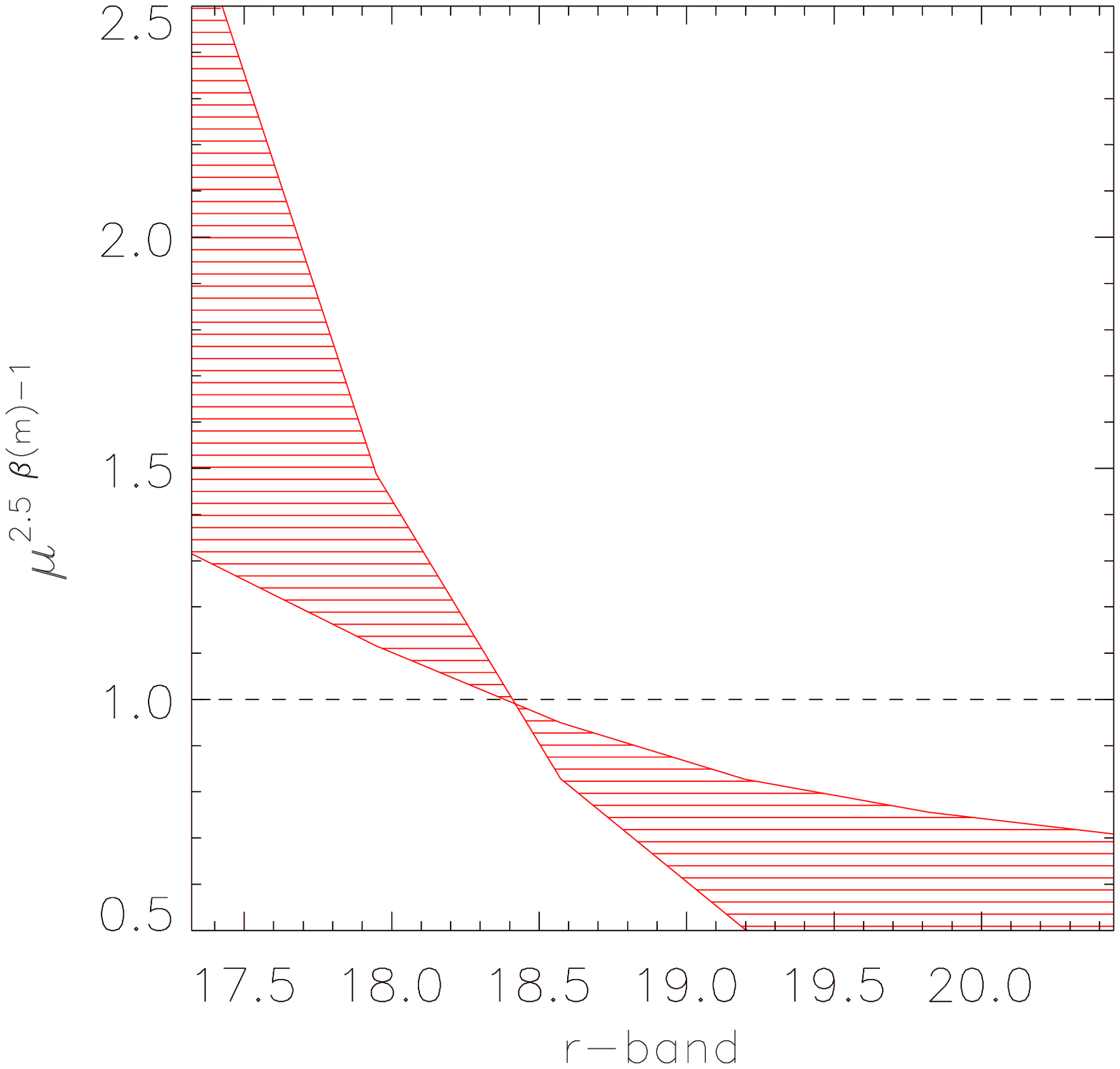}\hfill
  \caption{The value of the gravitational magnification bias
    $\mu^{2.5\beta(m)-1}$ is plotted as a function of magnitude, for the u-,
    b- and r-bands. The amplification $\mu$ depends on the lens properties and
    the coefficient $\beta$ is the logarithmic slope of the reference quasar
    number counts (see Fig. \ref{plot_nb_counts}).  As described by Eq.
    \ref{eq_mag} this quantity is proportional to the expected ratio between
    the number of quasars with and without absorbers as a function of
    magnitude.  We are considering here a \emph{simplified} model with
    amplifications ranging from $\mu=$1.07 to 1.3, as motivated in the text.
    The \emph{overall gradients} given by this magnification bias are
    comparable to the one observed from the data. Note that the amplitudes can
    not be compared given the need of normalizing the samples.}
\label{plot_alpha}
\end{figure*}

In order to \emph{illustrate} that the lensing explanation is in quantitative
agreement with our measurements we consider now a \emph{simplified} scenario and
compute the corresponding lensing effects.  
For estimating the amplification $\mu$ we assume the dark matter distribution
of the absorbers to be isothermal with a velocity dispersion between 100 and
200 $\mathrm{km\,s^{-1}}$ as it is expected for spiral galaxies. For a quasar
at $z=2$, an absorber at $z=1$ and an effective impact parameter of 10 kpc
(following the spatial distribution given by Steidel (1993) for DLAs) we find
an amplification $1.07\lesssim\mu\lesssim 1.3$ for $\Omega_m=0.3$ and
$\Omega_\Lambda=0.7$.  The impact parameter being a few times larger than the
Einstein radius of the lens, the by-pass effect becomes rather weak (see Fig.
3 in Smette et al. 1997) and can be neglected in the following.  Therefore the
total magnification bias is simply described by Eq.  \ref{eq_mag}.  Using the
values of $\beta(m)$ introduced above, we plot $\mu^{2.5\beta(m)-1}$ in Fig.
\ref{plot_alpha} for the three different bands.  The corresponding curves
describe the relative excess of quasars with an absorber and are therefore
directly related to the lower panels of Fig.  \ref{plot_results_tot}.
Our simplified model does not aim at reproducing the exact amplitude and shape
of expected lensing effects. It is only an indication of the magnification
bias behaviour. As we can in Fig. \ref{plot_alpha}, it shows that the overall
\emph{gradients} obtained from gravitational lensing are similar to the ones
measured from the data in the previous section (note that the amplitudes
cannot be compared in a straightforward manner contrary to the gradients).
This quantitative result strengthens the evidence towards gravitational
lensing being at the origin of the tilt measured in this study.  It also shows
how the signal found in section \ref{detection} is related to the average magnification
due to absorber halos and can therefore constrain their mass distribution.

More detailed calculations taking into account the redshift distributions of
the quasars and the absorbers as well as a distribution of impact parameters
and the inclusion of the by-pass effect are beyond the scope of this paper;
but they will be required in the future in order to estimate the expected
lensing effects more accurately and to infer some constraints from such
observations on the potential wells of these systems.

\subsection{Intrinsic Absorbers}
\label{intrinsic}

Despite the fact that intervening galaxies are clearly at the origin of some
absorption lines, a number of our absorber systems may actually be physically
associated to the quasars. Such associations have been confirmed when broad
absorption lines are seen in quasar spectra (Turnshek 1988, Weymann 1997) but
the situation is still unclear in the case of narrow lines (Borgeest \& Mehlert
1993).  Recently, Richards \e\ (1999, 2001) and Richards (2001) observed
correlations between quasar properties (magnitude, spectral index) and the
number of absorbers found in their spectra. They suggested that the presence
of absorbers might be related to the orientation of the quasars and thus gives
rise to some correlations with quasar magnitudes.

To investigate whether associations could be responsible for the magnitude
differences observed in section \ref{detection}, we have redone our analysis for
subsamples of absorbers with different escape velocities.  Indeed,
associations can no longer be at the origin of observed correlations when the
velocity difference between a quasar and a metal absorber is a substantial
fraction of the speed of light. 
%
\begin{figure}[h]
\begin{center}
  \includegraphics[width=0.9\hsize,height=7cm]{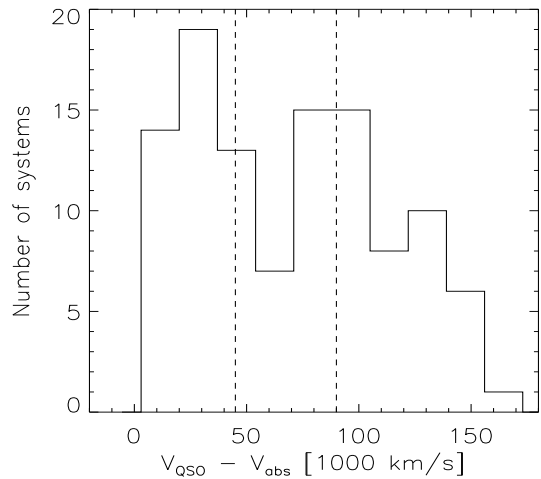}\hfill
  \caption{Distribution of absorber escape velocities.
    The vertical lines located at 45\,000 and 90\,000 km/s indicate the limits
    used for the two low and high escape-velocity subsamples.}
\label{plot_beta}
\end{center}
\end{figure}
Figure \ref{plot_beta} shows the distribution of the blueshifted absorber
velocities relative to the emission-line redshift of the QSOs, given by:
\begin{equation}
\beta=\frac{v}{c}=\frac{(1+z_\mathrm{QSO})^2-(1+z_\mathrm{abs})^2}
     {(1+z_\mathrm{QSO})^2+(1+z_\mathrm{abs})^2}\,.
\end{equation}
In order to see how the signal varies as a function of $\beta$ we analyse two
subsamples of the same size: one with escape velocities $v>90\times
10^3\,\kms$, i.e. for which associations can hardly be invoked, and a
corresponding low-escape velocity sample with $v<45\times 10^3\,\kms$. Each
sample has 39 systems. The results are summarized in Table
\ref{table_velocities} where we present the values of the gradients $\gamma$
and their significance, for the u-, b- and r-bands. The size of these
subsamples being reduced by almost a factor three compared to the measurement
performed in section \ref{detection}, it is expected to have significantly
lower detection levels.  Nethertheless we can observe that the signal can be
detected in each case at the $2\sigma$ level at least. Within the accuracy we
can reach, the $\gamma$ values do not show any specific trend with escape
velocity.

\begin{table}[ht]
\begin{center}

\begin{tabular}{ll}
  \begin{tabular}{l}
\hspace{-.3cm}Low escape velocity~~~~~$\,$\\
\hspace{-.3cm}$v<45\times 10^3$km/s\\
\hspace{-.3cm}39 QSO with abs.
    \end{tabular}&  \begin{tabular}{lcc}
                        \hline\hline
                        band    &gradient $\gamma$ &detection\\
                        \hline 
                        u &-0.41$\pm 0.20$  & 1.9$\sigma$\\
                        b &-0.88$\pm 0.27$  & 3.2$\sigma$\\
                        r &-0.74$\pm 0.25$  & 2.9$\sigma$\\
                        \hline\hline
                      \end{tabular}\\
~\\
\end{tabular}

\begin{tabular}{ll}
  \begin{tabular}{l}
\hspace{-.3cm}Medium escape velocity\\
\hspace{-.3cm}$v>90\times 10^3$km/s\\
\hspace{-.3cm}39 QSO with abs.
    \end{tabular}&  \begin{tabular}{lcc}
                        \hline\hline
                        band    &gradient $\gamma$ &detection\\
                        \hline 
                        u &-0.47$\pm 0.21$  & 2.0$\sigma$\\
                        b &-0.67$\pm 0.27$  & 2.2$\sigma$\\
                        r &-0.57$\pm 0.24$  & 2.2$\sigma$\\
                        \hline\hline
                      \end{tabular}\\
~\\
\end{tabular}

\end{center}
\caption{Values of the gradients $\gamma$ found for absorber subsamples of different
escape velocity ranges. At low escape velocities, either gravitational lensing
or physical associations might be able to give rise to negative $\gamma$
values, whereas only lensing seems to be able to be at the origin of the
effect found for the high-escape velocity sample.}
\label{table_velocities}
\end{table}

For the low-escape velocity sample the detection of the tilt could arise
either from correlations due to associated absorbers, as suggested by Richards
et al. (1999, 2001), or from gravitational lensing effects.  In this case,
each scenario is plausible.  However, the detection of the tilt for the
high-escape velocity sample can hardly be explained by associations since this
case deals with velocities higher than roughly one third of the light speed.
Therefore, for this subsample, we can fully attribute the excess of bright
(and/or deficit of faint) quasars with an absorber to gravitational lensing effects.\\

Since such a lensing signal contains valuable information about the
gravitational potential of the absorbers, it would be of great interest to
estimate the fraction of associated absorbers as a function of escape
velocity. This will allow us to maximise the number of quasars for which only
lensing induced correlations are expected and therefore improve the accuracy
of the magnification bias measurement.

\section{Conclusion}

Using the first release of the 2dF Quasar survey (2QZ), we have looked for the
magnification bias due to intervening absorption systems along the
line-of-sight of quasars. Contrarily to previous studies which searched
changes in quasar magnitudes due to presence of numerous but weak absorbers in
quasar spectra, we have focused on the strongest absorption systems.
\newline The 2QZ sample lists over 10$\,$000 quasars of which 1363 were visually
inspected by Outram \e\ (2001) for the compilation of a strong metal
absorber catalogue. Motivated by the idea that some of these systems may trace
galaxies, a \emph{magnification bias} is expected to modify the magnitude
distribution of the corresponding background quasars (Bartelmann \& Loeb 1995,
Pei 1995, Perna et al.  1997, Smette et al.  1997).  In order to measure such
an effect, we have carefully selected samples of quasars and absorbers
according to the following main steps:
\begin{itemize}
  
\item from the catalogue compiled by Outram et al. (2001) we selected the
  \MgIIFeII\ absorbers. Note that these systems have
  $1.3\lesssim\,$W$_0\,($\MgII\ doublet$)\lesssim 9.0$ \AA ,
  
\item for the corresponding observed equivalent widths, we have checked that
  the detection of these absorption lines is not biased with respect to the
  signal-to-noise of the spectra,

\item we define a reference population by bootstrapping the
population of quasars without absorber with a redshift
distribution identical to the one of quasars with an absorber.

\end{itemize}

By comparing the magnitude distributions of the resulting quasar populations,
we have showed that these are significantly tilted: an excess of bright
(and/or a deficit of faint) quasars with an absorber is detected at the 2.4,
3.7 and 4.4$\sigma$ levels in the u, b and r-bands.
Moreover a consistent signal is still detected (at the 3.3$\sigma$ in r-band)
if we use only the absorbers with the highest observed equivalent widths
(W$>7.5$\AA ), i.e. systems much less sensitive to biases in the detection
procedure.
Given the similar redshift and signal-to-noise properties imposed by our
sample selection, this magnitude difference is necessarily related to the
presence of the absorbers. This detection implies that number counts of
quasars with strong \MgIIFeII\ absorbers are substantially modified with
respect to quasars without such absorbers along their line-of-sight and
corrections have to be applied in order to recover the true underlying
properties of these objects.

We review different effects arising when a matter concentration intercepts the
light coming from a quasar: extinction and reddening, flux contributions from
the absorbing system, gravitational lensing and correlations due to physical
quasar-absorber associations. We argued that only the two latter explanations
can give rise to the trend we observe.
We have then redone our analysis on two subsamples of the same size having low
($v<45\times 10^3\,\kms$) and high ($v>90\times 10^3\,\kms$) quasar-absorber
velocity differences.  In each case we detected similar magnitude changes (at
the 1.9 to 3.2$\sigma$).  For the low-escape velocity sample, either
gravitational lensing or physical associations could be at the origin of the
observed correlation.  However for the high velocity sample, the only likely
explanation for the changes in the magnitude of quasars with an absorber is
gravitational lensing since, in this case, the velocity differences between
quasars and absorbers are greater than roughly one third of the light speed.
Being able to isolate and measure such a lensing signal is of great interest
since this magnification bias contains information about the absorber
gravitational potential.

Our analysis also shows that the changes in quasar magnitudes due to the
presence of absorbers tend to be stronger in redder bands. This trend
might be explained by extinction effects since they give rise to an effect
opposite to that of gravitational lensing, i.e. a relative excess of
\emph{faint} quasars with an absorber, which is expected to be stronger in
bluer bands. Isolating gravitational lensing from extinction effects would be
possible by using complete quasar samples in different bands.\\

The effects of the magnification bias depend on the slope of the source
number-counts as a function of magnitude and therefore on the characteristics
of a given survey.  In the case of 2QZ quasars, the corresponding number count
slopes are expected to give rise to a relative excess of bright quasars, in
agreement with our detection.  Moreover, we have shown that a
\emph{simplified} gravitational lensing model gives quantitative tilt
predictions comparable to the ones observed in the magnitude distribution of
quasars with absorbers.  More accurate measurements of this effect might
therefore give us interesting constraints on the absorber mass distribution.

It should be emphasized that such a lensing technique allows us to
greatly extend the usual redshift ranges probed by existing statistical
shear or magnification measurements (see Mellier 1999 for a review).
Indeed, given the need of background galaxies, these previous
measurements were restricted to z$_{\rm lens}\lesssim 1$, whereas the
use of quasars and absorbers allows us to easily probe lensing effects
at z$_{\rm lens}\sim$1--2.  Moreover the lenses are selected according to their
lensing optical depth as opposed to mass or luminosity.\\

We are currently undertaking a similar analysis using SDSS spectra.  Using a
different survey and facing different systematics will allow us to check the
significance of the present detection.  Besides the larger sample sizes that
Sloan will provide, the better spectrum quality will allow the detection of
various kinds of absorption lines. Therefore we will be able to extend the
analysis to different metal absorbers and see, via the magnification bias, how
they populate dark matter halos. This technique might therefore provide a
promising tool to get new constrains on the nature of high-redshift absorber
systems.

\section*{Acknowledgments}
We are grateful to Jacqueline Bergeron, Simon White and Matthias Bartelmann
for fruitful discussions, to Alain Smette for a careful reading of the
manuscript and to the referee, Donald York, for many valuable comments on an
earlier version of the manuscript. CP is supported by the Marie Curie program
of the European Community.  This work was supported in part by the TMR Network
``Gravitational Lensing: New Constraints on Cosmology and the Distribution of
Dark Matter'' of the EC under contract No.  ERBFMRX-CT97-0172.


\newpage
\appendix

\section{The sample of quasars with \MgIIFeII\ absorbers}

This appendix gives explicitly the list of quasars with absorbers we use in
our analysis. The initial sample comes from Outram et al. (2001). We refer the
reader to this paper for the details about the compilation of this catalogue,
as well as more information on each system.

Table \ref{table_summary} summarises the main steps we used in order to select
our sample of quasars with strong \MgIIFeII\ absorbers. A detailed list of the
corresponding objects is given in Table \ref{table_details}.  From the list of
quasars with \MgIIFeII absorbers given by Outram et al., we do not take into
account two systems in our analysis: the first one (flagged R1) is a quasar
not detected in the r-band. The second one (flagged R2) has an absorber escape
velocity smaller than $3\,000$ km/s.
\newline The A flag indicates the presence of two absorbers in the quasar
spectrum and the B flag is used when one of these two absorbers has
$z_{\rm{abs}} \approx z_{\rm{QSO}}$.

\begin{table}[ht]
\caption{Summary of the successive steps of the sample selection with
the corresponding number of quasars.}
\begin{center}
\begin{tabular}{lcc}
\hline \hline ~ & reference &quasars with\\ 
Selection criteria &quasars &absorber\\ 
\hline 
1. Initial catalogue from Outram et al.& 1135 & 129\\ 
~~~ (2001)\\ 
2. Selecting only quasars with \MgII\ & 1135 & 110\\
~~~ and \FeII\ systems\\
3. Excluding quasars not detected&1114 & 109 \\ 
~~~ in the r-band& & \\ 
4. Rejecting associated absorbers&1114 & 108 \\
~~~ with $\Delta v<3000$ km/s& & \\ 
\hline\hline
\end{tabular}
\end{center}
\label{table_summary}
\end{table}

\begin{table}[h]
\caption{Detailed list of the quasars with an absorber used in our analysis.}
\begin{center}

  \begin{tabular}{l|ccclc}
    \hline 
    \hline 
    Quasar          &$z_{\rm em}$ &S/N  &$z_{\rm abs}$   &W$_0$   &Flag\\
    \hline 

    J000534.0-290308         &2.347  &19         &1.168 &4.83 & \\
    J000811.6-310508         &1.683  &25         &0.715 &2.55 & \\
    J001123.8-292500         &1.280  &20         &0.605 &6.82 & \\
    J001233.1-292718         &1.565  &16         &0.913 &2.83 & \\
    J002832.4-271917         &1.622  &47         &0.753 &1.83 & \\
    J003142.9-292434         &1.586  &23         &0.930 &5.39 & \\
    J003533.7-291246         &1.492  &17         &1.457 &3.78 & \\
    J003843.9-301511         &1.319  &43         &0.979 &2.93 & \\
    J004406.3-302640         &2.203  &22         &1.042 &3.10 & \\
    J005628.5-290104         &1.809  &23         &1.409 &3.63 & \\
    J011102.0-284307         &1.479  &26         &1.156 &3.24 & \\
    J011720.9-295813         &1.646  &36         &0.793 &2.53 &R1\\
    J012012.8-301106         &1.195  &64         &0.684 &4.01 & \\
    J012315.6-293615         &1.423  &16         &1.113 &2.30 & \\
    J013032.6-285017         &1.670  &16         &1.516 &3.37 & \\
    J013356.8-292223         &2.222  &17         &0.838 &4.64 & \\
  \end{tabular}
\end{center}
\label{table_details}
\end{table}

\begin{table}
  \begin{tabular}{lccclc}
    \hline 
    \hline 
    Quasar          &$z_{\rm em}$ &S/N  &$z_{\rm abs}$   &W$_0$   &Flag\\
    \hline 
    J013659.8-294727         &1.319  &17         &1.295 &3.01 & \\
    J014729.4-272915         &1.697  &15         &0.811 &3.92 & \\
    J014844.9-302817         &1.109  &49         &0.867 &1.75 & \\
    J015550.0-283833         &0.946  &35         &0.677 &2.62 & \\
    J015553.8-302650         &1.512  &16         &1.316 &3.18 & \\
    J015647.9-283143         &0.919  &17         &0.884 &3.14 & \\
    J015929.7-310619         &1.275  &28         &1.079 &1.56 & \\
    J021134.8-293751         &0.786  &18         &0.616 &3.45 & \\
    J021826.9-292121         &2.469  &19         &1.205 &4.45 & \\
    J022215.6-273231         &1.724  &23         &0.611 &2.55 & \\
    J022620.4-285751         &2.171  &18         &1.022 &9.03 & \\
    J023212.9-291450         &1.835  &15         &1.212 &4.63 & A\\
    J024824.4-310944         &1.399  &26         &0.789 &4.91 & A, B\\
    J025259.6-321125         &1.954  &17         &1.735 &3.94$^a$ & \\
    J025608.0-311732         &1.255  &33         &0.973 &4.62 & A\\
    J025919.2-321650         &1.557  &16         &1.356 &3.15 & \\
    J030249.6-321600         &0.898  &48         &0.821 &4.54 & \\
    J030324.3-300734         &1.713  &42         &1.190 &2.95 & \\
    J030647.6-302021         &0.806  &21         &0.745 &4.21 & \\
    J030711.4-303935         &1.181  &25         &0.966 &2.81 & A, B\\
    J030718.5-302517         &0.992  &25         &0.711 &4.95 & \\
    J030944.7-285513         &2.117  &21         &0.931 &3.39 & \\
    J031255.0-281020         &0.954  &15         &0.955 &2.06 & R2\\ 
    J031309.2-280807         &1.435  &21         &0.950 &1.78 & \\
    J031426.9-301133         &2.071  &25         &1.128 &6.08 & A\\
    J095605.0-015037         &1.188  &20         &1.045 &3.04 & \\
    J095938.2-003501         &1.875  &27         &1.598 &4.31 & \\
    J101230.1-010743         &2.360  &17         &1.370 &2.83 & \\
    J101556.2-003506         &2.462  &17         &1.489 &2.29 & \\
    J101636.2-023422         &1.519  &18         &0.912 &2.90 & \\
    J101742.3+013216         &1.457  &18         &1.313 &1.72$^a$ & \\
    J102645.2-022101         &2.401  &23         &1.581 &1.31 & \\
    J105304.0-020114         &1.527  &16         &0.888 &5.76 & \\
    J105620.0-000852         &1.440  &22         &1.285 &1.70 & \\
    J110603.4+002207         &1.659  &36         &1.018 &2.08 & \\
    J110736.6+000328         &1.726  &51         &0.953 &2.65 & \\
    J114101.3+000825         &1.573  &17         &0.841 &3.93 & \\
    J115352.0-024609         &1.835  &20         &1.204 &2.90 & \\
    J115559.7-015420         &1.261  &19         &1.132 &3.75 & \\
    J120455.1+002640         &1.557  &16         &0.597 &3.92 & \\
    J120826.9-020531         &1.724  &20         &0.761 &5.31 & \\
    J120827.0-014524         &1.552  &15         &0.621 &3.13 & \\
    J120836.2-020727         &1.081  &27         &0.873 &1.97 & \\
    J122454.4-012753         &1.347  &20         &1.089 &2.14 & \\
    J125031.6+000216         &2.100  &20         &1.327 &2.61 & \\
    J125658.3-002123         &1.273  &28         &0.947 &3.62 & \\
    J130019.9+002641         &1.748  &17         &1.225 &5.92 & \\
    J130433.0-013916         &1.596  &19         &1.410 &8.05 & \\
    J130622.8-014541         &2.152  &16         &1.332 &3.67 & \\
    J133052.4+003219         &1.474  &52         &1.327 &1.89 & \\
    J134448.0-005257         &2.083  &18         &0.932 &5.44 & \\
    J135941.1-002016         &1.389  &30         &1.120 &2.97 & \\
    J140224.1+003001         &2.411  &24         &1.387 &2.85 & \\
    J140710.5-004915         &1.509  &16         &1.484 &3.49 & \\
    J141051.2+001546         &2.598  &16         &1.170 &4.74 & \\
    J144715.4-014836         &1.606  &18         &1.354 &2.02 & \\
    J214726.8-291017         &1.678  &35         &0.931 &1.74 & \\
    J214836.0-275854         &1.998  &55         &1.112 &1.40 & \\
    J215024.1-282508         &2.655  &30         &1.144 &1.88 & \\
  \end{tabular}
\end{table}

\begin{table}
  \begin{tabular}{lccclc}
    \hline 
    \hline 
    Quasar          &$z_{\rm em}$ &S/N  &$z_{\rm abs}$   &W$_0$   &Flag\\
    \hline 
    J215034.6-280520         &1.358  &35         &1.139 &1.58 & \\
    J215222.9-283549         &1.228  &24         &0.927 &2.59 & \\
    J215342.9-301413         &1.729  &31         &1.037 &1.54 & \\
    J215359.0-292108         &1.160  &22         &1.036 &3.33 & \\
    J215955.4-292909         &1.477  &23         &1.241 &6.24 & \\
    J220003.0-320156         &2.047  &16         &1.135 &3.20 & \\
    J220137.0-290743         &1.266  &24         &0.600 &3.81 & \\
    J220208.5-292422         &1.522  &47         &1.490 &2.98 & \\
    J220214.0-293039         &2.259  &78         &1.219 &3.39 & \\
    J220655.3-313621         &1.550  &15         &0.754 &4.60 & \\
    J221155.2-272427         &2.209  &35         &1.390 &2.76 & \\
    J221546.4-273441         &1.967  &20         &0.785 &2.86 & \\
    J222849.4-304735         &1.948  &33         &1.094 &3.86 & \\
    J223309.9-310617         &1.702  &47         &1.146 &2.58 & \\
    J224009.4-311420         &1.861  &29         &1.450 &2.04$^b$ & \\
    J225915.2-285458         &1.986  &18         &1.405 &4.57 & \\
    J230214.7-312139         &1.699  &21         &0.955 &1.98 & \\
    J230829.8-285651         &1.291  &43         &0.726 &3.94 & \\
    J230915.3-273509         &2.823  &18         &1.060 &4.66 & \\
    J231227.4-311814         &2.743  &20         &1.555 &2.95 & \\
    J231459.5-291146         &1.795  &39         &1.402 &3.12 & \\
    J232023.2-301506         &1.149  &17         &1.078 &3.50 & \\
    J232330.4-292123         &1.547  &17         &0.811 &3.70 & \\
    J232700.2-302637         &1.921  &38         &1.476 &6.41 & A\\
    J232914.9-301339         &1.494  &20         &1.294 &2.78 & \\
    J232942.3-302348         &1.829  &15         &1.581 &6.99 & \\
    J233940.1-312036         &2.611  &27         &1.444 &2.34 & \\
    J234321.6-304036         &1.956  &28         &1.052 &2.87 & \\
    J234400.8-293224         &1.517  &35         &0.851 &3.13 & \\
    J234405.7-295533         &1.705  &19         &1.359 &2.88 & \\
    J234527.5-311843         &2.065  &48         &0.828 &6.67 & \\
    J234550.4-313612         &1.649  &39         &1.138 &1.95 & \\
    J234753.0-304508         &1.659  &39         &1.421 &2.49 & \\
    J235714.9-273659         &1.732  &24         &0.814 &3.52 & \\
    J235722.1-303513         &1.910  &19         &1.309 &3.65 & \\
\hline    
\end{tabular}
\end{table}
\noindent
\scriptsize{$^a$ Rest equivalent width of \FeII\ $\lambda$=2600 \AA\ (\MgII\
  blended with sky lines)}\\
\scriptsize{$^b$ Rest equivalent width of \MgII\ $\lambda$=2796 \AA\ only.


\end{document}